\documentclass[journal,10pt,oneside,twocolumn]{IEEEtran}

\usepackage{cite}
\usepackage{amsmath}
\usepackage{amsfonts}
\usepackage{graphicx}
\usepackage{subfig}
\usepackage{xcolor}

\newcommand\hlbreakable[1]{\textcolor{black}{#1}}
\usepackage[displaymath]{lineno}

\begin{document}
\title{Cooperative Perception With Learning-Based V2V Communications}

\author{Chenguang~Liu,  Yunfei~Chen, {\em{Senior~Member,~IEEE}}, Jianjun~Chen,\\
Ryan~Payton, Michael~Riley, and~Shuang-Hua~Yang, {\em{Senior~Member,~IEEE}}

\thanks{This work was supported in part by Oracle Cloud credits and related resources provided by the Oracle for Research program, the National Natural Science Foundation of China (Grant No. 61873119, 92067109 and 62211530106), Shenzhen Science and Technology Program (Grant No. ZDSYS 20210623092007023 and GJHZ 20210705141808024), and the Educational Commission of Guangdong Province (Grant No. 2019KZDZX1018). (\it{Corresponding author: Shuang-Hua Yang.})}
\thanks{Chenguang Liu is with the School of Engineering, University of Warwick, Coventry, UK, CV4 7AL. {e-mail: chenguang.liu@warwick.ac.uk}}
\thanks{Yunfei Chen is with the Department of Engineering, University of Durham, Durham, UK, DH1 3LE. {e-mail: yunfei.chen@durham.ac.uk}}
\thanks{Jianjun Chen is with the Faculty of Engineering and Information Technology, University of Technology Sydney, Australia. {e-mail: jianjun.chen@student.uts.edu.au}}
\thanks{Ryan Payton and Michael Riley are with Oracle for Research, Oracle UK. {email: ryan.payton@oracle.com; michael.riley@oracle.com}}
\thanks{Shuang-Hua Yang is with Shenzhen Key laboratory of Safety and Security for Next Generation of Industrial Internet, Southern University of Science and Technology, Shenzhen, China, and also with Department of Computer Science, University of Reading, UK. {e-mail: yangsh@sustech.edu.cn}}
}
\markboth{Journal of \LaTeX\ Class Files,~Vol.~14, No.~8, August~2015}%
{Liu \MakeLowercase{\textit{et al.}}: Cooperative perception with learning-based V2V communications}

\maketitle

\begin{abstract}
Cooperative perception has been widely used in autonomous driving to alleviate the inherent limitation of single automated vehicle perception. To enable cooperation, vehicle-to-vehicle (V2V) communication plays an indispensable role. This work analyzes the performance of cooperative perception accounting for communications channel impairments. Different fusion methods and channel impairments are evaluated. \hlbreakable{A new late fusion scheme is proposed to leverage the robustness of intermediate features. In order to compress the data size incurred by cooperation, a convolution neural network-based autoencoder is adopted.} Numerical results demonstrate that intermediate fusion is more robust to channel impairments than early fusion and late fusion, when the SNR is greater than 0 dB. Also, the proposed fusion scheme outperforms the conventional late fusion using detection outputs, and autoencoder provides a good compromise between detection accuracy and bandwidth usage. 

\end{abstract}

\begin{IEEEkeywords}
Cooperative perception, machine learning, V2V communications.
\end{IEEEkeywords}

\IEEEpeerreviewmaketitle

\section{Introduction}
\IEEEPARstart{P}{erception} of the dynamic environment plays a vital role in autonomous driving. Due to the advancement of sensor technologies and deep learning algorithms, the performance of 3D detection for autonomous vehicles has been continuously improved with 3D scanners, such as light detection and ranging (LiDAR) \cite{8578570,s18103337}. LiDAR can sweep the surrounding by emitting and detecting the reflected laser light to obtain a 3D point cloud, which contains rich geometric, shape, and scale information. However, the detection performance of a single autonomous vehicle (AV) relies on the onboard sensors' physical capabilities, including resolution, detection range, and scan frequency. When the target objects are occluded or distantly located, a single automated vehicle cannot provide sufficient information to detect and range this object accurately. This performance degradation may consequently lead to severe driving safety issues. 

To alleviate the inherent limitation of a single AV perception, a cooperative perception system has been proposed, which consist of one ego car and multiple connected automated vehicles (CAVs) to leverage the information from multiple viewpoints. The CAVs can transmit their sensed information to the ego vehicle through vehicle-to-vehicle (V2V) communications so that the ego vehicle can aggregate the received information for best detection. Several works have been conducted on cooperative perception using different types of shared information, including raw point cloud data (\emph{i.e.} early fusion) \cite{8885377}, intermediate features (\emph{i.e.} intermediate fusion) \cite{10.1145/3318216.3363300,9812038, Wang2020V2VNetVC,li2021learning,pointpillars} and detection outputs of single CAVs (\emph{i.e.} late fusion) \cite{9228884}. However, although the transmission data size\cite{9812038,li2021learning}, time delay, and imperfect localization\cite{Wang2020V2VNetVC} have been considered in these works, none of them has considered communications channel impairments. In practice, V2V communications will suffer from the obstructed line of sight and channel distortion due to mobile antennas and multi-path fading. In a collaborative object detection system, these V2V channel impairments could result in severe performance degradation.

Previous works have neglected communication channel impairments. To address this, we first evaluate the detection performance of a cooperative perception system considering communications channel impairments for different fusion schemes and shared information. Based on this evaluation, we propose a new late fusion that utilizes intermediate features to leverage their robustness. Moreover, to reduce the data size incurred by fusion for cooperation and mitigate signal distortion, a convolution neural network (CNN)-based autoencoder is adopted. Previous works only used autoencoder to compress data but without channel impairment and signal distortion as studied in this work. Numerical results show this new fusion scheme outperforms the conventional late fusion, and the proposed use of autoencoder can balance the requirements on accuracy and data size with channel impairments.

\section{System model}
\begin{figure}[!htp]
  \centering
  \subfloat[]{\includegraphics[width=3in]{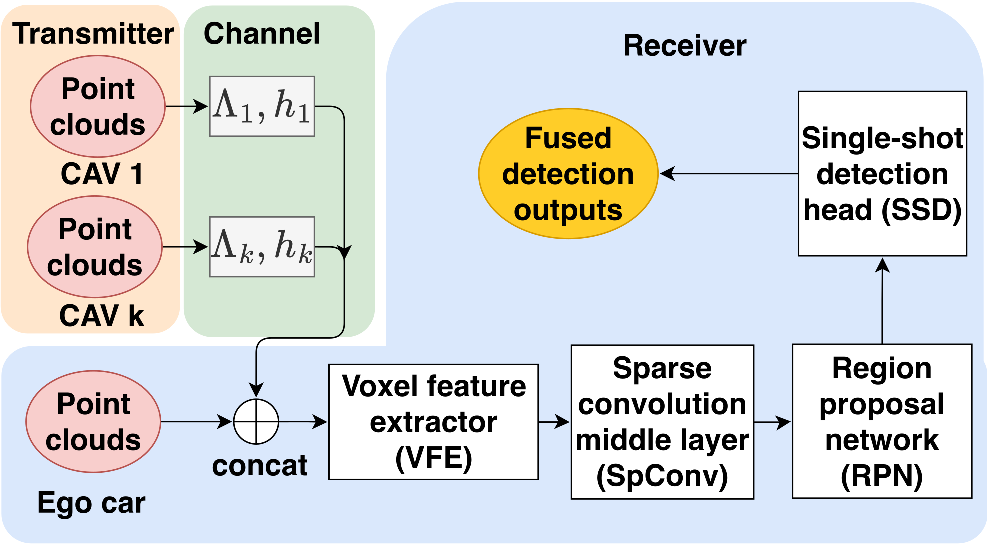}
  \label{1}}
  \hfil
  \subfloat[]{\includegraphics[width=3in]{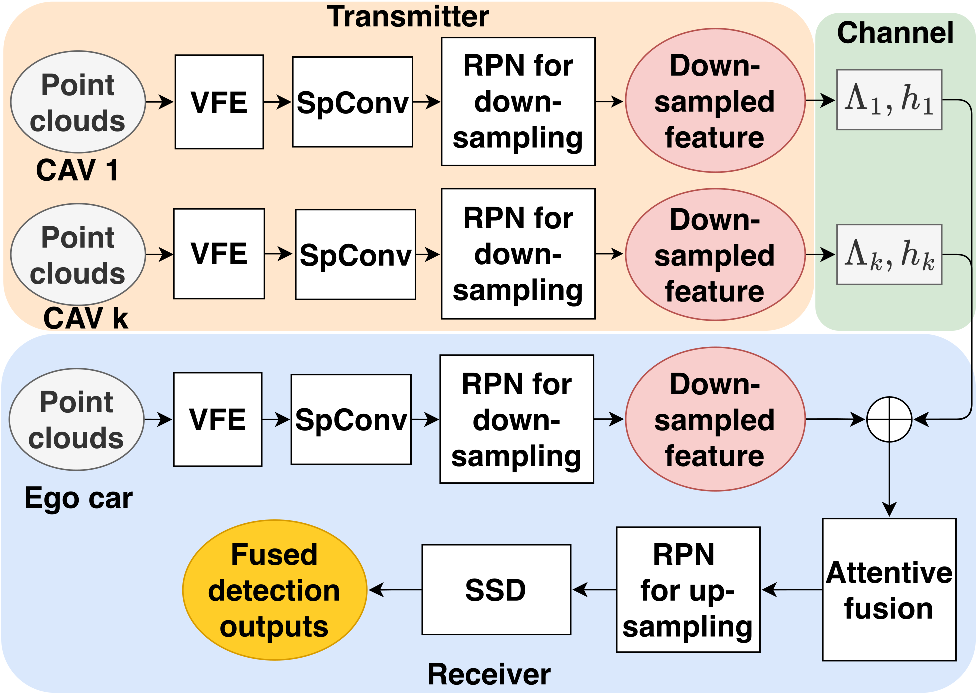}%
  \label{2}}
  \hfil
  \subfloat[]{\includegraphics[width=3.2in]{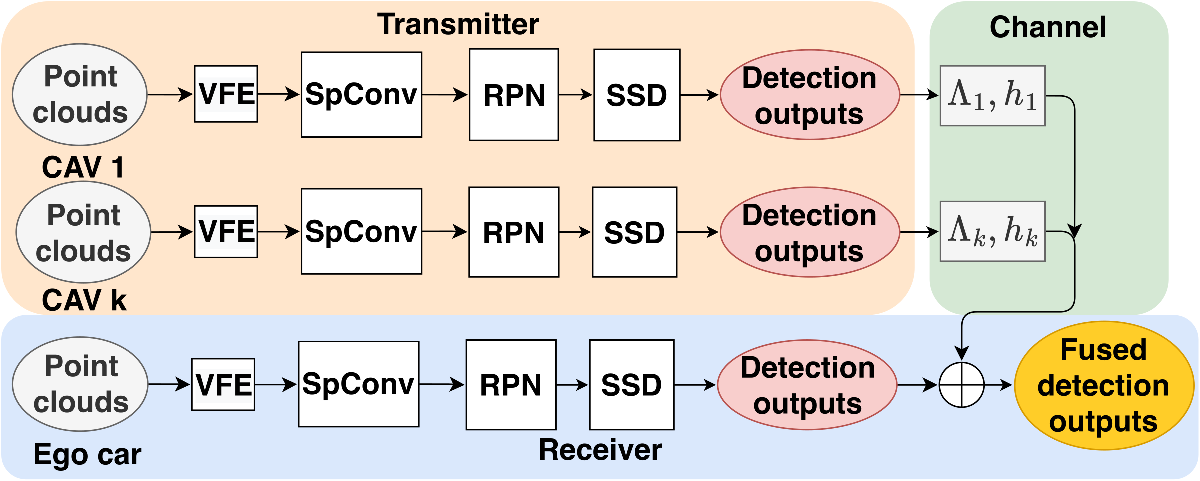}%
  \label{3}}
  \caption{Cooperative perception with V2V communications. (a) Early fusion. (b) Intermediate fusion. (c) Late fusion. }
  \label{system}
\end{figure}
\subsection{V2V communications model}
The V2V communications system for cooperative perception is shown in Fig. \ref{system}. Consider a single-input single-output (SISO) system with free-space path loss and Rician fading. When the $k$-th CAV transmits the information to the ego vehicle, the received signal is
\begin{equation}
    \mathbf{y}_k = \Lambda_k h_k\mathbf{x}_k + \mathbf{w}_k
\end{equation}
where $\Lambda_k=\sqrt{\frac{p_0}{d_k^n}}$ denotes the path loss with $p_0$ determined by antennas and channel characteristics, $d_k$ being the distance between transmitter and receiver and $n$ being the path loss factor, $\mathbf{y}_k \in \mathbb{C}^{L\times 1}$ and $\mathbf{x}_k \in \mathbb{C}^{L\times 1}$ denotes the complex-valued received signals and transmitted signals, respectively, $h_k$ denotes the Rician fading channel following $\mathcal{CN}(\mu,\sigma_h^2)$, $\mathbf{w}_k$ denotes the additive white Gaussian noise following $\mathcal{CN}(0,\sigma_w^2)$. Assume that the transmitted signals are recovered by zero-forcing detector with perfect channel state information (CSI).

To leverage the information from multiple CAVs, denote the shared information at the ego car as
\begin{equation}
    \mathbf{f} = F(\mathbf{x}_{ego},\mathbf{\hat{x}}_1,\mathbf{\hat{x}}_2,...,\mathbf{\hat{x}}_K ) 
\end{equation}
where $F(\cdot)$ denotes the fusion algorithm to aggregate the shared information, $\mathbf{x}_{ego}$ denotes the information sensed at the ego vehicle itself and $\mathbf{\hat{x}}_k$ denotes the recovered sensed information from $\mathbf{y}_k$ for the $k$-th CAV, $k=1,2,\cdots,K$. 

\subsection{Detection backbone}
In this work, the same backbone algorithm as SECOND \cite{s18103337} is used for cooperative perception. SECOND is a deep learning based approach, which inherits the end-to-end trainable structure of VoxelNet\cite{8578570} and adopts sparsely embedded convolutional layers to improve the efficiency of object detection.

The SECOND detector has three components: a voxel feature extractor (VFE), a sparse convolution middle layer, and a region proposal network (RPN). Firstly, raw point clouds are iteratively converted into voxel representation by assigning the points to the corresponding voxels. Then, a voxel-wise feature encoding layer is applied to the points in each voxel to extract their point-wise features. Subsequently, a sparse convolutional layer is applied to learn the 3D voxel features and reshape them to 2D image-like data. Specifically, sparse convolutional layers only apply the convolution to the non-zero inputs of the sparse embeddings, without processing all inputs. This could greatly reduce the computation cost and improve the efficiency when many of inputs are zero. Similar to the residual structure, RPN conducts two levels of downsampling and upsampling for the input features after sparse 3D convolution and then concatenates the outputs of each level into a feature map. Downsampling is performed by multiple layers of 2D convolution following batch normalization and ReLU activation, while upsampling utilizes one layer of deconvolution to restore the data size. Finally, a single-shot detector (SSD) is used to output the classification results of objects and the regression results of box localization. Instead of using separate networks in region proposals and objects, SSD generates object detection results with a single feedforward passing through the network, which is computationally more efficient.

\subsection{End-to-end training}
End-to-end training is adopted for cooperative perception with V2V communications, as the end-to-end trainable structure of SECOND allows any part of the information to be transmitted through the learning-based communications system. Since CAVs have different coordinates, the raw point clouds and detection outputs need to be normalized for each axis to follow the distribution with mean 0 and variance 1 for transmission. In this case, the learning-based communication and the detection algorithm can be trained together using the same loss function as in SECOND \cite{s18103337}. Signal distortion during communication can also be considered in training.

\section{Cooperative perception with learning-based V2V communications}
\subsection{Conventional fusion schemes}
As shown in Fig. \ref{system}, the conventional cooperative perception systems can be categorized into three types: early fusion, intermediate fusion and late fusion, which share the raw point clouds, intermediate features and detection outputs, respectively. The raw point cloud has the most information, but its transmission requires the largest bandwidth. The detection output consumes the least bandwidth but has the least information. Intermediate features are extracted from the raw point cloud to reduce the amount of the information transmitted. 

At the receiver side, early fusion simply concatenates the received point clouds to aggregate the information, since the raw point clouds are unordered and transformation invariant. Before concatenation, pre-processing is applied by cropping and filtering them according to the detection range to save the cost of computation. For intermediate fusion, such as attentive fusion \cite{9812038}, it adopts the downsampled features before deconvolution. Then, these features are processed by an attention-based neural network to obtain higher-level representations iteratively based on the number of residual layers. Unlike early fusion and intermediate fusion that require collaborative learning at the receiver, late fusion only uses the detection outputs from all CAVs. This simplifies the aggregation of the information from CAVs without offline learning. In late fusion, the detection outputs are selected by non-maximum suppression (NMS) and filtered by the detection range to obtain the final outputs.

These three systems have been proposed in the literature without considering wireless channel impairments. However, the raw data, intermediate features and detection outputs may have different robustness against channel impairments. Thus, this work will analyze their performances considering impairments.
\subsection{New late fusion using convolution features}
\begin{figure}[!ht]
    \centering
    \includegraphics[width=3.4in]{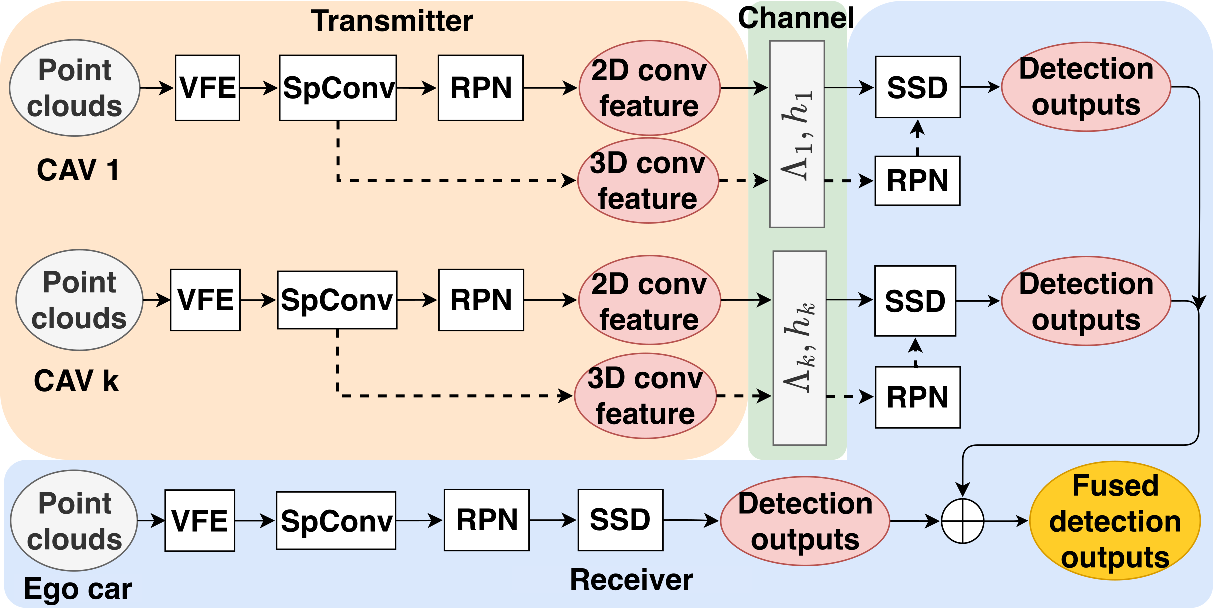}
    \caption{The cooperative perception using convolution features in late fusion.}
    \label{interlate}
\end{figure}
Based on the levels of convolution, the intermediate features can be the downsampled feature, the 3D convolution feature and the 2D convolution feature. The 3D and 2D convolution features are generated by the 3D SpConv and RPN, respectively, while the downsampled feature is by the downsampling convolution before deconvolution in RPN.

In the conventional scheme in Fig. \subref{2}, only the downsampled feature is used. In this work, new late fusion schemes using the 3D and 2D convolution features are proposed, as in Fig. \ref{interlate}. The reason is that intermediate convolution feature has rich information about the raw point clouds, and late fusion does not need offline training. This could provide robustness and computation efficiency for the ego vehicle. However, this requires the ego vehicle to correctly generate the detection outputs from received convolution features for each CAVs. Therefore, an autoencoder is applied as encoder-decoder to compress the convolution features at the transmitter and recover the desired information at the receiver. 

\section{Numerical results and discussion}
In this section, the effects of channel impairments are first examined. OPV2V\cite{9812038} is used as the training and evaluation dataset. Attentive fusion\cite{9812038} is used for the intermediate fusion. The training uses the Adam optimizer with a learning rate of 0.002, and the number of epochs is 60 with a batch size of 2. Consider Rician fading with Rician $K$ factor of 1 and additive white Gaussian noise with signal-to-noise ratio (SNR) from -10 to 30 dB. Average precision (AP) is adopted to measure the detection accuracy by calculating precision and recall values at different thresholds of intersection over union (IoU) and averaging the precision according to the recall values to obtain the AP score. Precision is the percentage of number of true positive out of the total detections, while recall refers to the percentage of number of true positive out of the total number of ground truth objects. The computation uses the Oracle Cloud Infrastructure with NVIDIA Tesla V100.
\begin{figure}[!h] 
  \centering
  \includegraphics[width=3.1in]{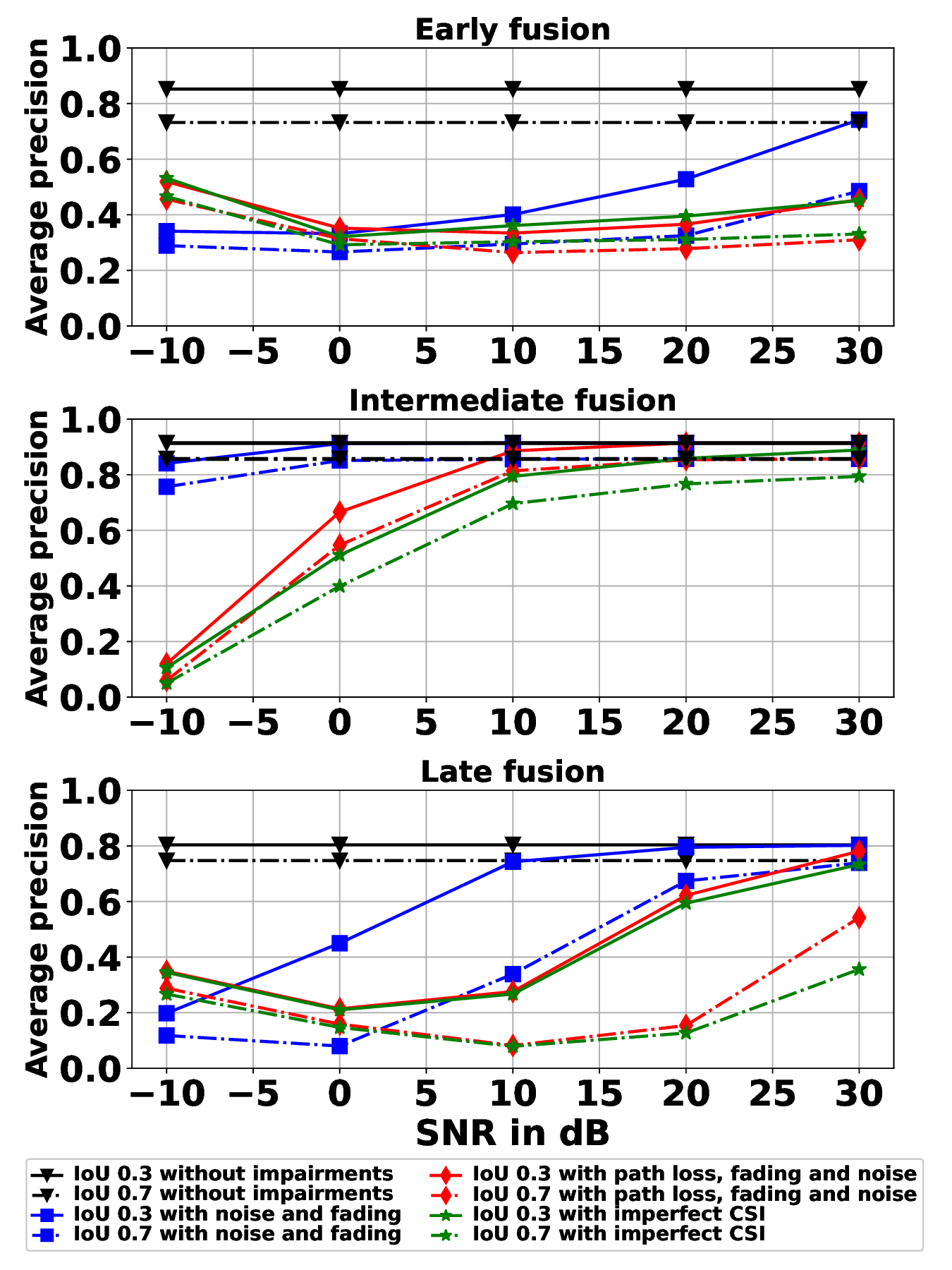}
  \caption{Average precision of cooperative perception for different fusion schemes.}
  \label{different_fusion_results}
\end{figure}
\subsection{Effects of channel impairment}\label{impairment}
Fig. \ref{different_fusion_results} shows the performance of cooperative perception for different fusion methods. For communications with noise and fading, the accuracy of early fusion has a stable increase from 30\% to 50\% for IoU = 0.7 and from 35\% to over 70\% for IoU = 0.3, when the SNR increases from -10 to 30 dB. For late fusion, the accuracy ranges from 20\% to 80\% for IoU = 0.3 and from 10\% to 75\% for IoU = 0.7, which outperforms early fusion when the SNR is larger than 10 dB. However, intermediate fusion has very stable accuracy over 76\% for IoU=0.7 and 80\% for IoU=0.3 for all SNRs. Thus, intermediate fusion has robustness to noise and fading. 

For communications with path loss, fading and noise, intermediate fusion has similar accuracy to the case without impairments when the SNR is greater than 10 dB. However, the accuracy drops from above 80\% to about 10\% when the SNR decreases from 10 to -10 dB. Different from the intermediate fusion, the accuracy of early fusion and late fusion increases as the SNR decreases from 10 to -10 dB. This is because the unreliable received raw point clouds and detection results are filtered out due to large distortion beyond the physical detection range. Thus the ego vehicle relies more on its own measurement when SNR is below 10 dB. Therefore, using more reliable raw point clouds and detection results could lead to an accuracy improvement. In addition to the path loss, a Gaussian disturbance with a mean 0 and variance 0.1 is added to the CSI to simulate the imperfect CSI. This causes around 10\% performance degradation to the intermediate fusion while it has limited effects on early fusion and late fusion when SNR is less than 20 dB.
\subsection{Effects of path loss factor} \label{pathloss}
\begin{figure}[!h]
  \centering
  \includegraphics[width=2.8in]{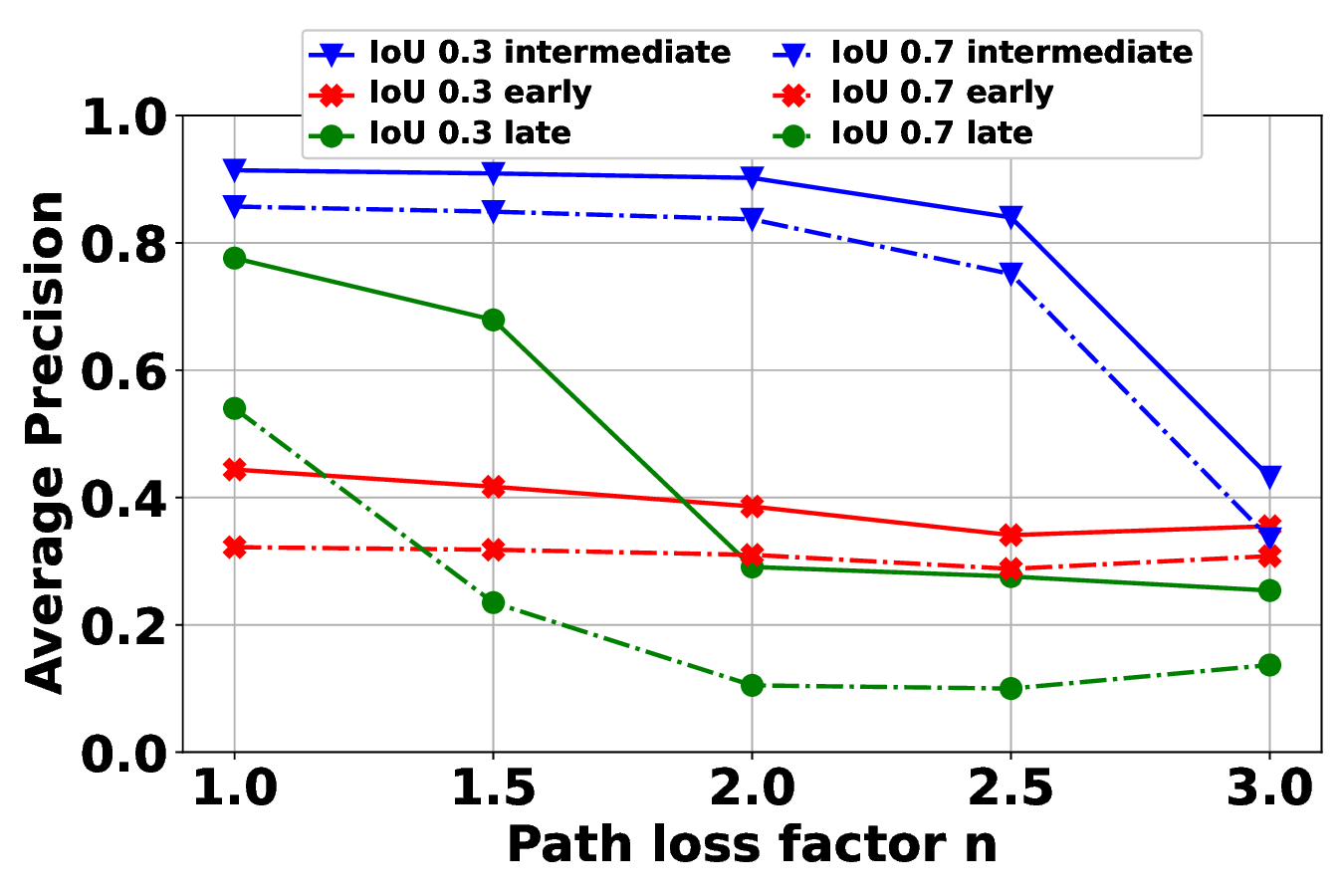}
  \caption{Average precision for different path loss factors.}
  \label{path_loss}
\end{figure}
Fig. \ref{path_loss} shows how the performance varies with the path loss factor. For early fusion, it is barely affected by the path loss factor but its accuracy remains low at around 30\% and 50\% for IoU = 0.3 and 0.7, respectively. Late fusion degrades from over 75\% and 50\% to around 10\% and 30\%, as the path loss factor increases from 1 to 2 and remains low. Intermediate fusion has high accuracy of above 80\% until $n=2.5$ before a sharp drop to about 40\% when $n=3$. Thus, transmitting intermediate features is the best choice when $n\leq2.5$. However, early fusion has a stable performance regardless of the path loss factor and a similar accuracy to intermediate fusion for IoU = 0.7 when $n=3$. Considering the large bandwidth required for transmitting raw point clouds, transmitting intermediate features is more cost-efficient.

\subsection{New late fusion using 3D and 2D features} 
\begin{figure}[!h]
  \centering
  \includegraphics[width=3in]{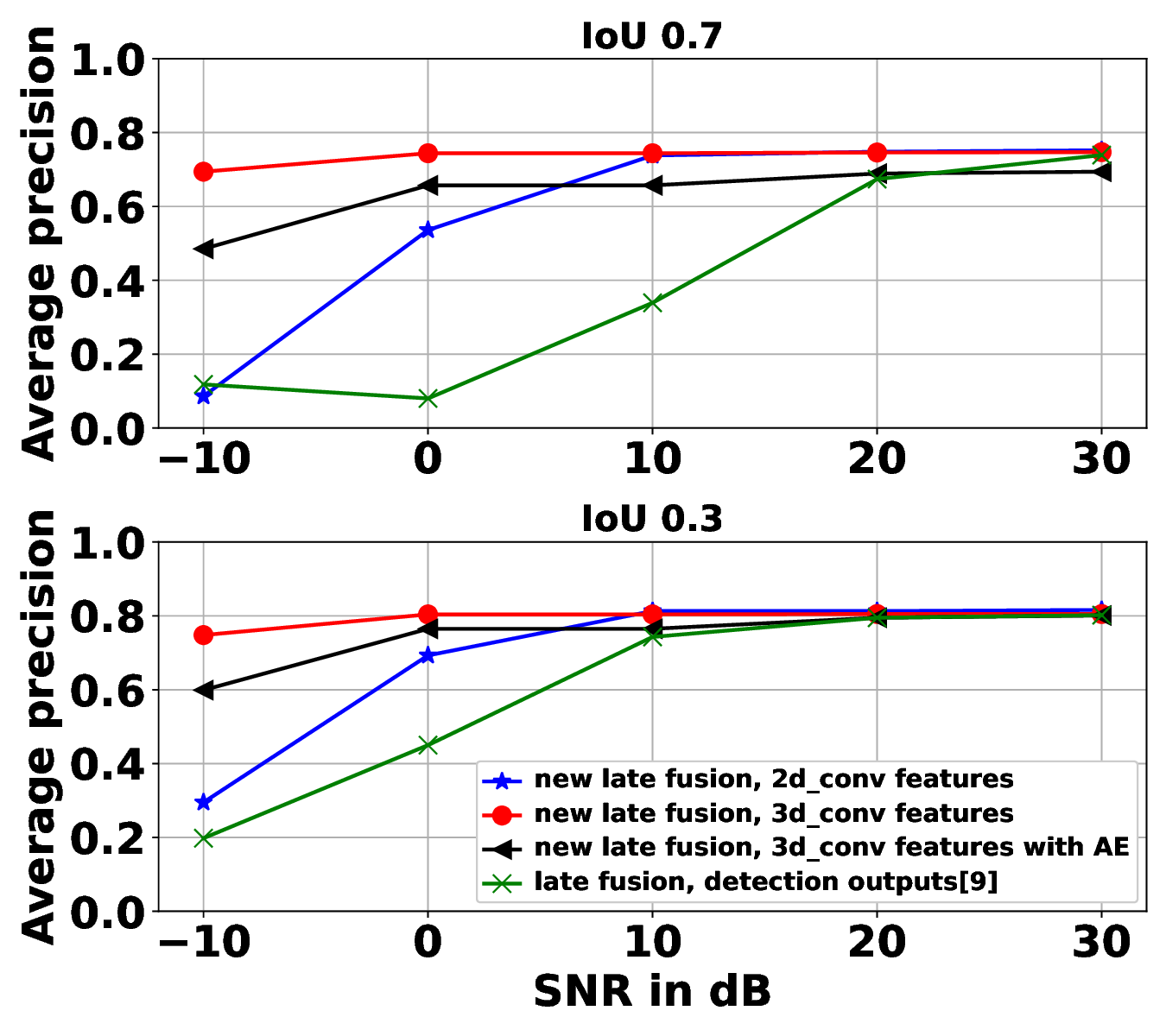}
  \caption{Average precision using different intermediate features in the late fusion.}
  \label{intermediate_late}
\end{figure}
Fig. \ref{intermediate_late} demonstrates the performances of late fusion using different convolution features. A CNN-based autoencoder is also used as the encoder-decoder to compress the 3D features, which has 64 times of data compression for the sparse features before transmission. First, new schemes using 2D and 3D convolution features perform better than the conventional late fusion using detection outputs, as the intermediate features are more robust than the detection results as transmitted information. Furthermore, 3D convolution features can achieve over 70\% accuracy even when the SNR is -10 dB. It outperforms 2D convolution features when the SNR is less than 0 dB. Comparing it with the case using autoencoder, one sees that the autoencoder loses about 5\% accuracy when the SNR is greater than 10 dB, and this loss increases to around 20\% when the SNR decreases to -10 dB. This performance loss can be considered as the tradeoff for the 64x data compression of the transmitted information. However, it is still better than 2D convolution features and detection outputs, especially when the SNR is between -10 and 0 dB. Thus, autoencoder is beneficial to effectively compress the 3D convolution features without losing too much accuracy. 

\section{Conclusion}
This work has studied a V2V communications model for cooperative perception and proposed the use of new intermediate features in late fusion with a CNN-based autoencoder. Different fusion methods and transmitted information have been evaluated for different SNRs and path loss factors. Numerical results have shown that intermediate fusion have better robustness against fading, noise, and path loss than the early fusion and late fusion. Also, using intermediate convolution features in the late fusion can significantly outperform the conventional late fusion using detection outputs. Numerical results have also shown that autoencoder is capable of compressing data with acceptable performance loss with channel impairments.

\ifCLASSOPTIONcaptionsoff
  \newpage
\fi




\end{document}